\documentclass[5p,sort&compress]{elsarticle}
\usepackage{amssymb}
\usepackage{amsmath}
\usepackage{graphicx}
\usepackage[draft]{hyperref}
\usepackage[utf8]{inputenc}
\usepackage{color}
\usepackage{enumitem}

\renewcommand{\L}{\mathcal{L}}

\renewcommand{\d}{\partial}

\renewcommand{\a}{\alpha}

\begin{document}

\title{Dilaton-assisted generation of the Fermi scale from the Planck scale }

\author[epfl]{Andrey Shkerin}
\ead{Andrey.Shkerin@epfl.ch}

\address[epfl]{
 Institute of Physics, Laboratory for Particle Physics and Cosmology,
\'{E}cole Polytechnique F\'{e}d\'{e}rale de Lausanne, CH-1015 Lausanne, 
Switzerland}

\begin{abstract}

In scale-invariant theories of gravity the Planck mass $M_P$, which appears due to spontaneous symmetry breaking, can be the only scale at the classical level. It was argued that the second scale can be generated by a quantum non-perturbative gravitational effect. The new scale, associated with the Higgs vacuum expectation value, can be orders of magnitude below $M_P$, leading to the hierarchy between the Fermi and the Planck scales. We study a theory in which the non-perturbative effect is sensitive both to the physics at energy scales as high as $M_P$ and to the low-energy, Standard Model physics. This makes it possible to constrain the mechanism from experiment. We find that the crucial ingredients of the mechanism are non-minimal coupling of the scalar fields to gravity, the approximate Weyl invariance at high energies, and the metastability of the low-energy vacuum.

\end{abstract}


\maketitle

\section{Introduction and setup}
\label{sec:Intro}

In the Standard Model (SM), the Higgs mass $m_H$ is the only scale at the classical level. Setting it to zero enhances the symmetry of the model by making it conformally invariant (CI). It is natural, therefore, to start from a theory without the classical electroweak (EW) symmetry breaking, and generate $m_H$ due to some quantum effect \cite{tHooft:1979rat,Bardeen:1995kv} (see also \cite{Wetterich:2019qzx}). One possible way to do this is via the quantum conformal anomaly \cite{Coleman:1970je}. It is known that radiative corrections can change the shape of the tree-level Higgs potential in such a way that a minimum displaced from the origin appears \cite{Coleman:1973jx,Weinberg:1978ym}. However, within the SM, this scenario predicts the values of the Higgs and the top masses being far from those observed experimentally \cite{Linde:1975sw,Weinberg:1976pe,Linde:1977mm,Witten:1980ez}. To overcome this issue, one can look for different extensions of the SM by new bosonic degrees of freedom; see, e.g., \cite{Loebbert:2018xsd} and references therein.

If the physics beyond the Standard Model (BSM) contains new heavy degrees of freedom, it is, in general, a source of large perturbative corrections to the Higgs mass; for overviews of the problem see \cite{Giudice:2008bi,Giudice:2013nak}. As long as heavy particles are coupled to the Higgs field, they are expected to produce contributions to $m_H$ that drive the latter towards the mass scale of those particles. One of the possibilities to cancel these large contributions is to make the BSM physics manifest itself close to the EW scale. The examples include supersymmetry, composite Higgs models, and extra dimensions \cite{Feng:2013pwa,Bellazzini:2014yua,ArkaniHamed:1998rs}. Given the absence of signatures of new physics at the TeV scale \cite{Giudice:2013nak}, the theories extending the SM by introducing mass thresholds right above the EW scale become confronted with the problem of fine-tuning, unless a special mechanism stabilizes $m_H$ against large corrections (see, e.g., \cite{Dvali:2003br,Dvali:2004tma,Graham:2015cka,Giudice:2016yja}). In theories with no heavy mass thresholds it was argued that quadratic divergences, which arise when one uses a mass-dependent regularization scheme, are not physical and do not pose a problem \cite{Vissani:1997ys,Shaposhnikov:2007nj,Farina:2013mla}.

It is customary to cope with puzzling facts about the SM physics (such as the quadratic divergences in $m_H$) by invoking new phenomena at nearby energy scales. An alternative point of view is to suggest that the low-energy observables (such as the Higgs vacuum expectation value (vev) $v\approx 246$ GeV \cite{Tanabashi:2018oca}) are sensitive rather to the physics \textit{much} above the SM energy domain; see \cite{Shaposhnikov:2018xkv,Shaposhnikov:2018jag} and references therein. In the CI SM framework and having assumed no new heavy degrees of freedom, a plausible candidate for such physics is quantum gravity. 

In this paper, we discuss a possible mechanism of generation of the Higgs vev, in which gravity plays a crucial role. In the (nearly) scale-invariant perturbation theory based on dimensional regularization, gravitational corrections to the Higgs potential are suppressed by the (reduced) Planck mass $M_P=2.435\cdot 10^{18}$ GeV and are numerically small at the weak scale \cite{tHooft:1972tcz,tHooft:1974toh,Shaposhnikov:2007nj}. Hence, the mechanism must work due to some non-perturbative gravitational effect. This idea was expressed and implemented in \cite{Shaposhnikov:2018xkv,Shaposhnikov:2018jag}. There it was argued that gravity is indeed capable of generating a new mass scale associated with the Higgs vev. Under certain assumptions about the high-energy behavior of the theory, the new scale can be many orders of magnitude below $M_P$, thus reproducing the observed hierarchy.

The idea behind the mechanism is discussed in \cite{Shaposhnikov:2018jag}, and we briefly outline it here. For simplicity, consider a theory of one real scalar field $\varphi$ coupled to dynamical gravity. The coupling can be non-minimal: for example, in \cite{Shaposhnikov:2018xkv} the models containing the term $\propto R\varphi^2$ were studied.\footnote{A motivation behind studying this type of model lies in their application to cosmology and inflation \cite{Bezrukov:2008ut, GarciaBellido:2008ab,Bezrukov:2014ipa,Kubo:2018kho}.} If we switch to the Euclidean signature,\footnote{See sec. \ref{ssec:Lagr} for the discussion regarding the Euclidean formulation of a theory.} then the spatially homogeneous time-independent vev of $\varphi$ can be evaluated as follows:
\begin{equation}\label{PathIntegral}
\langle\varphi\rangle\sim\int D\varphi Dg_{\mu\nu} \: e^{\log\varphi(0)-S_E} \; ,
\end{equation}
where $g_{\mu\nu}$ is the metric field, and $S_E$ is the Euclidean action of the theory. We now want to evaluate the path integral in eq. (\ref{PathIntegral}) via saddle points of the functional
\begin{equation}\label{ActionEnh}
\mathcal{B}=-\log\varphi(0)+S_E \; .
\end{equation}
The legitimacy of such an evaluation is discussed in \cite{Shaposhnikov:2018jag}, and we will repeat the arguments below in this paper. The saddle points of $\mathcal{B}$ are instantons satisfying a boundary condition provided by the instantaneous source term $-\log\varphi(0)$. Thus, the non-perturbative contribution to the vev of $\varphi$ is provided by a suitable instanton configuration. In the semiclassical approximation
\begin{equation}\label{Hierarchy}
\langle\varphi\rangle\sim M_Pe^{-B} \; ,
\end{equation}
where $B$ is the instanton value of $\mathcal{B}$. 

In the general case, the form of the source term in eq. (\ref{ActionEnh}) depends on the content of the theory. For example, in \cite{Shaposhnikov:2018jag} SI models of gravity with two scalar degrees of freedom were studied, and the source was shown to be a logarithm of a certain polynomial function of the scalar fields.

According to eq. (\ref{Hierarchy}), in order to reproduce the observed ratio between the Fermi and the Planck scales, 
\begin{equation}\label{TheHierarchy}
v/M_P\sim 10^{-16} \; ,
\end{equation}
one must have $B=\log M_P/v\approx 37$. The instanton action $B$ is a function of various couplings present in the theory.\footnote{However, in the models studied in \cite{Shaposhnikov:2018xkv,Shaposhnikov:2018jag}, the main behavior of $B$ is controlled by a single parameter --- the non-minimal coupling of a scalar field to gravity. } For some of their values it turns out that $|B|\lesssim 1$, and the semiclassical approximation made in going from eq. (\ref{PathIntegral}) to eq. (\ref{Hierarchy}) is not valid. A possible interpretation of this fact is that in this case quantum gravity effects are strong, they rise $\langle\varphi\rangle$ up to $M_P$, and no new scale appears. In other regions of the parameter space it is possible to obtain $B\gg 1$. Now one can say that the non-perturbative gravitational effects are suppressed, and the hierarchy of scales emerges. 


All instanton solutions studied in \cite{Shaposhnikov:2018xkv,Shaposhnikov:2018jag} turn out to be insensitive to the behavior of the theory at energy scales much below $M_P$. In general, the instanton action $B$ can be split into the sum of two terms. The first of them, $B_{HE}$, represents the contribution from the core region of the instanton, which probes the UV structure of the theory. The second term, $B_{LE}$, is the contribution from the tail of the instanton, which is determined by low-energy physics. The scale separating the two domains can be identified with a cutoff of the low-energy effective theory.\footnote{Note that, because of the presence of gravity, the theory under consideration is non-renormalizable. } Then,
\begin{equation}
B=B_{HE}+B_{LE} \; .
\end{equation}
The insensitivity of the solution to the low-energy physics and, in particular, to the SM parameters yields $B\approx B_{HE}$. On the one hand, it is intriguing to conclude that the observed hierarchy between the weak and the Planck scales can result purely from features of the theory in the strong-gravity regime. On the other hand, a certain drawback of this conclusion is that the properties that can be measured in experiment (for example, the shape of the effective Higgs potential in the CI SM) have no impact on the instanton mechanism and do not put any constraints on it. 

In this paper, we study a SI theory of gravity, in which both the high-energy (the core) and the low-energy (the tail) parts of the instanton configuration contribute significantly to the total instanton action. The two regions of energy scales at which the action is saturated can be separated by many orders of magnitude. This happens due to a peculiar behavior of the instanton profile, namely, due to the fact that the profile is not a monotonic function of the distance from the center of the configuration. 

The short-distance behavior of the instanton shares many of its properties with the solutions studied previously in the context of the hierarchy problem \cite{Shaposhnikov:2018xkv,Shaposhnikov:2018jag}. We use similar arguments and employ the same type of UV operators to eliminate the divergence of the solution at the source point and to generate the large high-energy contribution $B_{HE}$. As for the tail of the instanton, we see that its shape closely resembles another Euclidean solution --- a bounce saturating the tunneling action \cite{Coleman:1977py,Coleman:1980aw}. In fact, the similarity between the two types of solutions goes much farther, as it turns out that they have the same existence conditions. In particular, it is necessary for the low-energy vacuum identified with the classical ground state of the theory to be the false vacuum state.

The intimate relation between the instanton and the bounce also implies that $B_{LE}\approx B_{bounce}$, where $B_{bounce}$ is the tunneling action. In a realistic setting this means that $B_{LE}=\mathcal{O}(10^3)$ \cite{Degrassi:2012ry,Buttazzo:2013uya,Andreassen:2017rzq}. Hence, in order to reproduce eq. (\ref{TheHierarchy}), $B_{HE}$ must be negative and the difference between the absolute values of $B_{LE}$ and $B_{HE}$ must be around $1\%$. Thus, a fine-tuning is required between the low-energy and the high-energy domains. Below we will see that this requirement severely constrains the parameter space of the theory. 

Let us set up the framework. Most of the paper deals with a SI model of gravity with two scalar degrees of freedom. The scale symmetry helps to protect the scalar field vev against large perturbative corrections;\footnote{Recall that we work under the assumption that there are no heavy degrees of freedom above the EW scale.} see, e.g.,  \cite{Shaposhnikov:2008xi}. Hence, it provides a suitable framework to attack the problem of hierarchy of scales with non-perturbative tools. One of the scalars (the dilaton) develops a classical vev associated with the Planck mass. Another field is associated with the Higgs degree of freedom, and its vev remains zero at the classical level. By supplementing the model with the rest of the SM content, one obtains the theory whose low-energy limit must reproduce the CI SM and general relativity. 

The choice of the model is motivated by its simple structure and by the possibility of disentangling the dilaton and the Higgs parts of the Lagrangian. At low energies, the model is an effective theory with the cutoff scale determined by $M_P$ and coupling constants. When the cutoff is approached, the low-energy Lagrangian must be supplemented with the set of higher-dimensional operators. Following the above arguments, we require that the scale symmetry be preserved in the UV regime of the theory, and that no heavy mass thresholds interfere between the weak and the Planck scales. The requirements constrain the set of irrelevant operators to be added at high energies, but their variety remains large, as it follows, e.g., from the Horndeski construction \cite{Horndeski:1974wa}. Our strategy is to probe a particular set of operators whose effect on the instanton results in the desired value of $B_{HE}$. In other words, we treat the Lagrangian of the model as relevant for the mechanism part of the classical Lagrangian of the fundamental theory.\footnote{The exception is the scalar field potential: the latter must be RG improved for the effect to appear; see, e.g., \cite{Andreassen:2017rzq} for the discussion of this observation in the context of EW vacuum decay in the SM.} We do not argue that the theory under consideration is consistent with a UV complete theory of gravity which is yet to appear. However, eq. (\ref{TheHierarchy}) can be viewed as an argument in favor of those properties of the fundamental theory that are relevant for the successful implementation of the mechanism.

The paper is organized as follows. In sec. \ref{sec:Model} we introduce the model, and explain its main features and its connection to phenomenology. In sec. \ref{sec:BI} Euclidean classical solutions arising in the model are studied. Namely, we describe the tunneling solution and the singular instanton. Similarities and differences between them are highlighted. In sec. \ref{sec:Hierarchy} we discuss the conditions under which the instanton mechanism of generating the hierarchy of scales works successfully and eq. (\ref{TheHierarchy}) is reproduced. Section \ref{sec:DiscConcl} contains the discussion of our findings and the conclusion.

\section{The model}
\label{sec:Model}

\subsection{The Lagrangian}
\label{ssec:Lagr}

In this and the next sections we focus on a particular model, in which we implement the instanton mechanism of generating the hierarchy of scales outlined in sec. \ref{sec:Intro}. We also demonstrate the sensitivity of this mechanism to the low-energy physics that shapes the instanton profile far from its core region. The model closely resembles the SI theories of gravity with two scalar fields studied in \cite{Shaposhnikov:2018jag}; yet, as we will see, the Euclidean solutions obtained here are qualitatively different from their counterparts considered previously.

We call the two scalar fields $\chi$ and $h$ the dilaton and the Higgs field, correspondingly. This agrees with \cite{Wetterich:1987fm,Shaposhnikov:2008xb,GarciaBellido:2011de,Bezrukov:2012hx}, where the Higgs-dilaton theory was studied. The global scale symmetry of the dilaton sector is broken spontaneously by the dilaton vev associated with the Planck scale. Our goal is to show that the semiclassical gravitational effect breaks the scale invariance of the Higgs sector, by generating the Higgs vev, and that the ratio of the two symmetry-breaking scales can be exponentially small.

To address the actual hierarchy problem (\ref{TheHierarchy}), it is necessary to supplement the model with the rest of the SM content. As long as we are interested in the leading-order contribution to the Higgs vev from the instanton built of $\chi$, $h$ and the metric fields, the other degrees of freedom can be ignored; we will briefly comment on their inclusion in sec. \ref{sec:Hierarchy}. Finally, for the sake of simplicity we work with the Euclidean formulation of the model. In doing so, one should be aware of possible issues with the high-energy limit of Euclidean quantum gravity \cite{Gibbons:1977zz}. We assume that the correct formulation of gravity in the UV regime resolves those issues one way or another.

The Euclidean Lagrangian of the model takes the form\footnote{The Lagrangian must be supplemented with a suitable boundary term. The latter, however, is not important for our analysis, and we will omit it \cite{Shaposhnikov:2018jag}.}
\begin{equation}\label{L_J}
\L_E=\L_\chi+\L_h \; ,
\end{equation}
where the dilaton Lagrangian is given by
\begin{equation}\label{L_chi}
\dfrac{\L_\chi}{\sqrt{g}}=-\dfrac{1}{2}\xi_\chi\chi^2R-\dfrac{1}{2}(\d\chi)^2+\delta\dfrac{(\d\chi)^4}{\chi^4} \; ,
\end{equation}
and the Higgs Lagrangian is
\begin{equation}\label{L_h}
\dfrac{\L_h}{\sqrt{g}}=-\dfrac{1}{2}\xi_h h^2R+\dfrac{1}{2}(\d h)^2+\dfrac{1}{4}\lambda h^4 \; .
\end{equation}
Let us comment on the structure of the model. First of all, both $\chi$ and $h$ are coupled non-minimally to gravity with the positive couplings $\xi_\chi$ and $\xi_h$ accordingly. Also, there is no direct coupling between the dilaton and the Higgs sectors. The signs of the non-minimal coupling terms are determined by the standard Euclidean continuation prescription \cite{Coleman:1980aw}. On the contrary, the sign of the dilaton kinetic term $(\d\chi)^2$ is chosen opposite to the standard sign of the Higgs kinetic term $(\d h)^2$. In general, this means that the total kinetic term of the scalar fields is not positive definite, and one should carefully determine the region of the parameter and field spaces where it remains positive definite. As we will see shortly, it is possible to avoid instabilities in fluctuations above the classical vacuum of the model. As for the Euclidean solutions, they will stay confined in the region of validity of the Lagrangian (\ref{L_J}). 

The choice of the sign of the dilaton kinetic term is the crucial feature of the model. If one switches it to the standard sign, one restores the theory whose close analogs were studied before in the context of the hierarchy problem \cite{Shaposhnikov:2018jag}, as well as in the light of their phenomenological implications \cite{GarciaBellido:2011de,Bezrukov:2012hx}. Because of this similarity between the two types of models, their Euclidean solutions have many features in common, but we also point out some remarkable differences between them.

As for the four-derivative term in eq. (\ref{L_chi}), its role is to regularize the divergence of the instanton solution and to provide a finite contribution $B_{HE}$ from its core region. After $\chi$ develops the classical vev, $\langle\chi\rangle=\chi_{vac}$, this term becomes suppressed by $\chi_{vac}^4$; therefore it controls the high-energy behavior of the model. We choose this particular SI higher-dimensional operator because of its particularly simple form; in sec. \ref{sec:DiscConcl} we will comment on a possible generalization of the UV part of the dilaton sector.

The Lagrangian (\ref{L_J}) is invariant under global scale transformations
\begin{equation}\label{SI_J}
g_{\mu\nu}\rightarrow\sigma^{-2}g_{\mu\nu} \; , ~~~ h\rightarrow\sigma h\; , ~~~ \chi\rightarrow\sigma\chi \; .
\end{equation}
The classical vacuum of the model provides one classical scale associated with the Planck mass:
\begin{equation}\label{GroundState_J}
\chi_{vac}\equiv\dfrac{M_P}{\sqrt{\xi_\chi}} \; , ~~~ h_{vac}=0 \; .
\end{equation}
Note that Lagrangian (\ref{L_J}) is not the most general Lagrangian consistent with the scale symmetry. From the point of view of the low-energy theory built above the vacuum (\ref{GroundState_J}), the quartic terms 
\begin{equation}\label{AddTerms}
\propto\chi^2 h^2 \; , ~~~ \propto\chi^4
\end{equation}
are also allowed. The first of these terms gives rise to the Higgs mass, and we set it to zero at the classical level. Then, all perturbative corrections to this term are suppressed by the Planck mass and, hence, are negligible at low energies. The next term, the dilaton quartic coupling term, gives rise to the cosmological constant. The latter is tiny in a realistic setting and can be neglected; see sec. \ref{ssec:Remarks} and \cite{Shaposhnikov:2018jag} for more details. 

It is worth noting that the terms in (\ref{AddTerms}) are not invariant under the constant shifts of the dilaton $\chi\rightarrow\chi+$const. Keeping them small is natural according to the 't Hooft definition \cite{tHooft:1979rat}, since in their absence (and in the limit $\xi_\chi\rightarrow 0$), the dilaton shift symmetry is restored \cite{Shaposhnikov:2008xi}.\footnote{ See also \cite{Karananas:2016grc} for the reasoning about a mechanism that could forbid the terms (\ref{AddTerms}). } However, the hierarchy between the couplings, which is necessary to reproduce the observed ratio between $m_H$ and the cosmological constant \cite{GarciaBellido:2011de}, is not respected by loop corrections, and the cosmological constant problem persists.

\subsection{Polar field coordinates}
\label{ssec:PolCoord}

To analyze classical solutions arising in the model (\ref{L_J}) and to understand the behavior of the scalar degrees of freedom, it is convenient to diagonalize the kinetic terms of the scalar fields. To this end, we perform a Weyl rescaling of the metric, which allows us to eliminate the non-minimal coupling, followed by the field redefinition:\footnote{The question of equivalence of theories related to each other by a Weyl rescaling of the metric is discussed in \cite{Falls:2018olk}. }
\begin{equation}\label{NewCoord}
\begin{split}
& g_{\mu\nu}=\Omega^{-2}\tilde{g}_{\mu\nu} \; , ~~ \Omega^2=\dfrac{\xi_\chi\chi^2+\xi_hh^2}{M_P^2} \; , \\
& \chi=\dfrac{M_P\cos\theta}{\sqrt{-1+6\xi_\chi}}\: e^{\rho/M_P} \; , \\
& h=\dfrac{M_P\sin\theta}{\sqrt{1+6\xi_h}}\: e^{\rho/M_P} \; .
\end{split}
\end{equation}
The fields $\rho$ and $\theta$ can be thought of as polar coordinates on the plane spanned by the vectors $\chi\sqrt{-1+6\xi_\chi}$ and $h\sqrt{1+6\xi_h}$. Note that the radial field $\rho$ is exponentiated in eqs. (\ref{NewCoord}). Thanks to the exponential mapping, the kinetic term of the scalar fields becomes canonical (up to constant multipliers) in the high-energy regime of the model. When evaluating the vev $\langle h\rangle$ in the path integral approach, the exponential term $\exp(\rho/M_P)$ is added to the Euclidean action of the model \cite{Shaposhnikov:2018jag}, resulting in the functional of the form (\ref{ActionEnh}) whose saddle points are the instanton solutions we are looking for.

In the new variables, the Lagrangian becomes
\begin{equation}\label{L_E}
\begin{split}
\dfrac{\L_E}{\sqrt{\tilde{g}}}=-\dfrac{1}{2}M_P^2\tilde{R} & +\dfrac{1}{2a(\theta)}(\tilde{\d}\rho)^2+\dfrac{b(\theta)}{2}(\tilde{\d}\theta)^2 \\
& +\tilde{V}(\theta)+\delta\dfrac{(\tilde{\d}\rho)^4}{M_P^4}+...
\end{split}
\end{equation}
where $\tilde{\d}$ means that contraction of the derivatives is performed with the new metric $\tilde{g}^{\mu\nu}$, and dots stand for the terms containing the powers of $\sin\theta\cdot\d\theta$, whose precise form will not be important in what follows. The various functions appearing in eq. (\ref{L_E}) are defined as follows:\footnote{Apart from the signs in $b(\theta)$ and $\zeta$, these expressions are the same as the ones obtained in \cite{GarciaBellido:2011de}.}
\begin{equation}\label{VariousF}
\begin{split}
& a(\theta)=\dfrac{1}{6+1/\xi_h}(\sin^2\theta+\zeta\cos^2\theta) \; , \\
& b(\theta)=\dfrac{M_P^2\zeta}{\xi_\chi}\dfrac{-\tan^2\theta+\xi_\chi/\xi_h}{\cos^2\theta(\tan^2\theta+\zeta)^2} \; , \\
& \tilde{V}(\theta)=\dfrac{\lambda M_P^4}{4\xi_h^2}\dfrac{1}{(1+\zeta \cot^2\theta)^2} \; , \\
& \zeta= \dfrac{(1+6\xi_h)\xi_\chi}{(-1+6\xi_\chi)\xi_h} \; .
\end{split}
\end{equation}

The action of the scale transformations (\ref{SI_J}) on the new field variables is given by
\begin{equation}\label{SI_E}
\tilde{g}_{\mu\nu}\rightarrow\tilde{g}_{\mu\nu} \; , ~~~  \rho\rightarrow\rho+\text{const.}  \; , ~~~ \theta\rightarrow\theta \; .
\end{equation}
To be invariant under the constant shifts of the field $\rho$, the Lagrangian (\ref{L_E}) must contain it only derivatively. The classical ground state (\ref{GroundState_J}) is written as
\begin{equation}\label{GroundState_E}
\rho_{vac}=\dfrac{M_P}{2}\log\left(\dfrac{-1+6\xi_\chi}{\xi_\chi}\right) \; , ~~~ \theta_{vac}=0 \; .
\end{equation}

From eq. (\ref{L_E}) it is easy to obtain the conditions under which the kinetic term of the scalar fields is positive definite. We take the following condition:
\begin{equation}\label{PosBounds}
\xi_\chi>\dfrac{1}{6} \; , ~~~ \xi_h>0 \; , ~~~ \theta<\theta_{max}=\arctan\sqrt{\dfrac{\xi_\chi}{\xi_h}} \; .
\end{equation}
According to the definition of $\theta$, the third of the above conditions sets an upper bound on the ratio $h/\chi$, above which the model must be modified to prevent the appearance of instabilities. The instanton solutions studied below all satisfy the condition $0\leqslant\theta<\theta_{max}$, and the range of validity of the Lagrangian (\ref{L_J}) is enough for our purposes. Note also that fluctuations $\delta\chi$, $\delta h$ of the scalar fields above the ground state are healthy as long as
\begin{equation}
\delta h<M_P\sqrt{\dfrac{-1+6\xi_\chi}{\xi_h(1+6\xi_h)}}\left(1+\dfrac{\delta\chi}{\chi_{vac}}\right) \; .
\end{equation}

In what follows, we allow the Higgs quartic coupling $\lambda$ to be the function of the angular variable, $\lambda=\lambda(\theta)$. The field dependence of the coupling mimics its RG running once extra degrees of freedom are added into the model, which are coupled to the Higgs field. It is necessary to take into account the running of $\lambda$, since the behavior of the instanton solution and its very existence depend strongly on the shape of the (effective) Higgs potential $\tilde{V}(\theta)$. The dependence of the normalization point on $\theta$, and not on $\rho$, follows from working in the SI renormalization scheme \cite{Englert:1976ep,Shaposhnikov:2008xi,Gretsch:2013ooa}; see sec. \ref{sec:Hierarchy}.

\section{The bounce and the instanton}
\label{sec:BI}

\subsection{General remarks}
\label{ssec:Remarks}

In this section, we study two distinct types of Euclidean classical configurations arising in the model (\ref{L_J}) written in the form (\ref{L_E}). The first is the bounce --- the regular solution that interpolates between the regions of the false and true vacua \cite{Coleman:1977py,Coleman:1980aw}. The bounce exists whenever the function $\lambda(\theta)$ becomes negative at some values of its argument. The potential $\tilde{V}(\theta)$ then becomes negative as well, its minimum at $\theta=\theta_{vac}=0$ is not global, and the ground state (\ref{GroundState_E}) represents the false vacuum state. Studying the bounce is important since, as we will see, the singular instanton solution --- the second type of classical configurations considered here --- inherits many of its properties.

Below we restrict our analysis to the $O(4)$-symmetric configurations. For the bounce, it is believed that the spherically symmetric solution saturates the tunneling action, although the proof is known only for a scalar field theory in flat space background \cite{Coleman:1977th,Blum:2016ipp}.\footnote{Note, however, that even in flat space, examples are known where a less symmetric configuration dominates the action \cite{Schmid:2007dm}.} For the instanton solution, one can note that the source term in the action (\ref{ActionEnh}) preserves the spherical symmetry of the model; hence it is natural to assume that the $O(4)$-symmetric solution provides a dominant contribution to the instanton action. We choose the following ansatz for the metric field $\tilde{g}_{\mu\nu}$:
\begin{equation}\label{Ansatz}
d\tilde{s}^2=f^2(r)dr^2+r^2d\Omega_3^2 \; ,
\end{equation}
where $d\Omega_3$ is the line element on a unit 3-sphere.

All solutions are required to approach the ground state (\ref{GroundState_E}) at large distances:
\begin{equation}\label{BC_Inf}
f^2\rightarrow 1\; , ~~~ \theta\rightarrow 0 \; , ~~~ \rho\rightarrow\rho_{vac} \; , ~~~ r\rightarrow\infty \; .
\end{equation}
Note that self-consistency requires us to look for configurations approaching the actual ground state with non-zero $\theta_{vac}$, which is obtained by taking into account the instanton contributions to the expectation values of $\chi$ and $h$. However, the difference between the actual solutions and the ones satisfying eqs. (\ref{BC_Inf}) is revealed at the distances $r\gtrsim v^{-1}$, where it cannot produce any noticeable effect. Note also that in our model the ground state is flat. It is possible to make it curved by introducing the term $\propto\chi^4$ into the dilaton Lagrangian (\ref{L_chi}).\footnote{Another way to introduce the cosmological constant is to replace general relativity by unimodular gravity; see \cite{Shaposhnikov:2008xb,Blas:2011ac} and references therein. } However, since in a phenomenologically interesting case the characteristic length of the background geometry significantly exceeds the characteristic sizes of the bounce and the instanton, switching to the ground state with non-zero curvature will not affect any of our results.

In explicit calculations below we will use the toy function $\lambda(\theta)$ that becomes negative well below $\theta_{max}$ for fixed $\xi_\chi$, $\xi_h\lesssim 1$. We choose $\lambda(\theta)$ so that the tunneling action is of the same order of magnitude as in the SM, $B_{bounce}=\mathcal{O}(10^3)$ \cite{Degrassi:2012ry,Buttazzo:2013uya,Andreassen:2017rzq}. One of our main results is that the low-energy contribution to the instanton action, $B_{LE}$, virtually coincides with $B_{bounce}$.

\subsection{The bounce}
\label{ssec:Bounce}

\begin{figure*}[t]
	\begin{center}
		\begin{minipage}[h]{0.49\linewidth}
			\center{\includegraphics[scale=0.65]{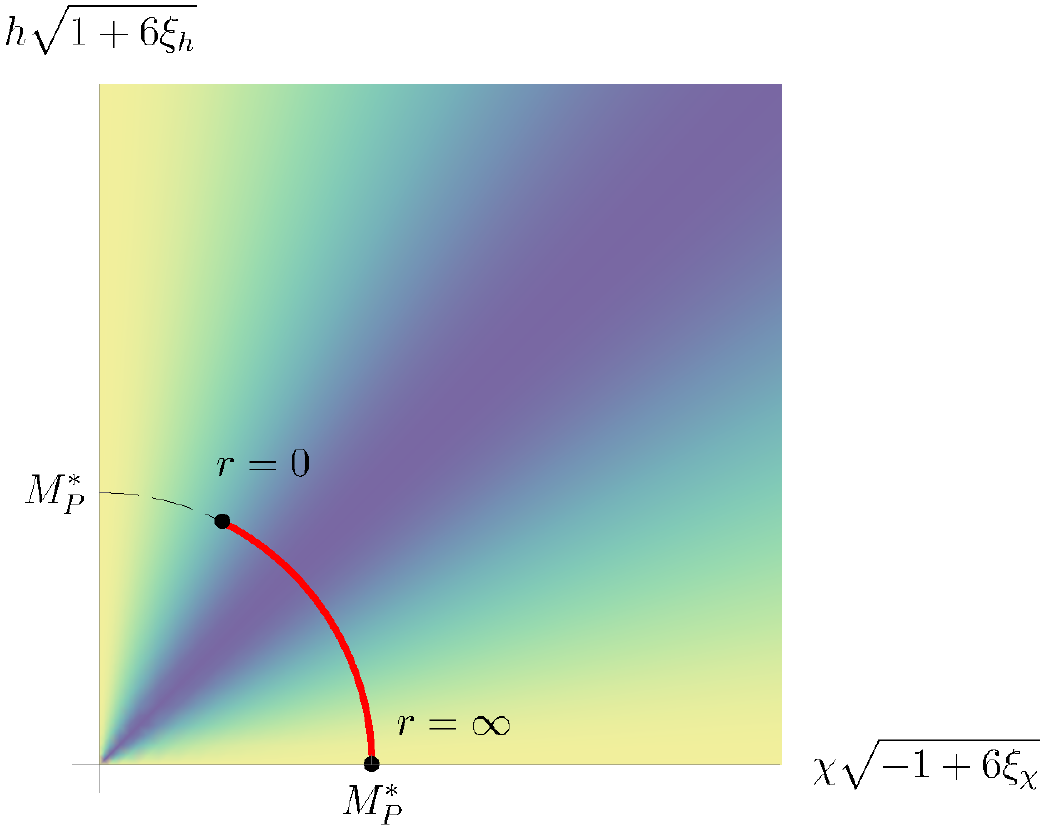} \\ (a)}
		\end{minipage}
		\begin{minipage}[h]{0.49\linewidth}
			\center{\includegraphics[scale=0.65]{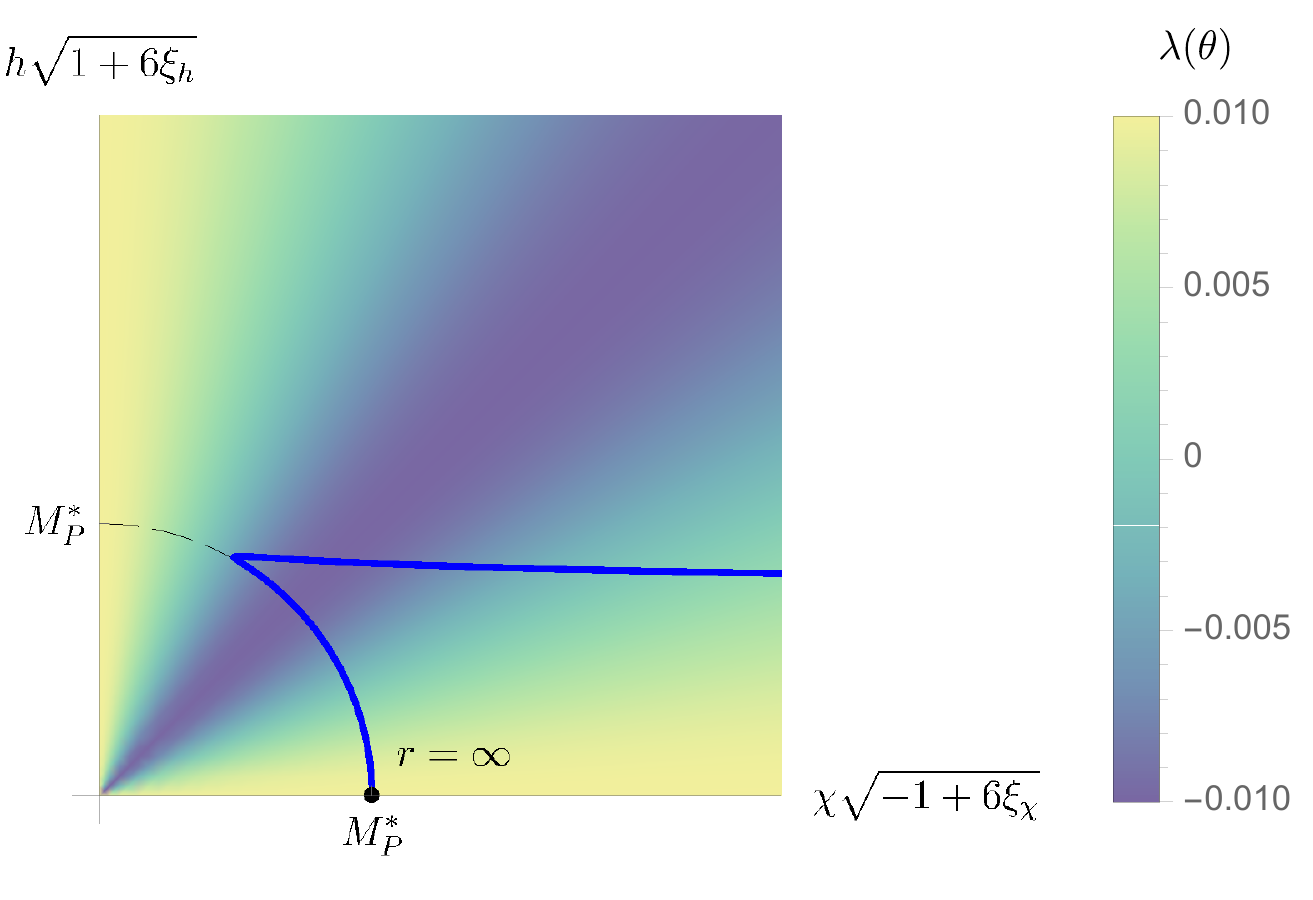} \\ (b)}
		\end{minipage}
		\caption{ The classical solutions in the model (\ref{L_J}) satisfying the boundary conditions (\ref{BC_Inf}) at infinity. We take $\lambda(\theta)=0.01\cos 4\theta$, $\xi_\chi=0.2$, $\xi_h=0.02$. \textit{Left panel:} The solid line represents the bounce, the dashed line is a circle of radius $M_P^*$ (see eqs. (\ref{b_eq}) and  (\ref{Mps})). \textit{Right panel:} The solid line represents the singular instanton solution. Its trajectory, if drawn from $r=\infty$, follows the path of the bounce until the turning point, after which $\theta$ starts decreasing according to the asymptotics (\ref{As_theta_zero2}), while $\rho$ starts growing according to eq. (\ref{As_rho_zero}).  }
		\label{fig:BI}
	\end{center}
\end{figure*}

Regularity of the bounce imposes the following boundary conditions at the origin:
\begin{equation}\label{BC_b_zero}
f^2(0)=1\; , ~~~ \theta'(0)=\rho'(0)=0 \; .
\end{equation}
Together with eqs. (\ref{BC_Inf}), they select a unique solution of the Euclidean equations of motion. The detailed analysis of the tunneling solution in a model similar to (\ref{L_J}) was performed in \cite{Shkerin:2016ssc}, where the question of stability of the EW vacuum in the Higgs-dilaton theory was addressed. Here we repeat the main steps of this analysis. First, let us switch the four-derivative term off for the moment, $\delta=0$. Then, applying the ansatz (\ref{Ansatz}) to the equation of motion of the field $\rho$, one finds
\begin{equation}\label{Eq_rho}
\dfrac{\rho'r^3}{fa(\theta)}=C
\end{equation}
with $C$ an arbitrary constant. Inspecting the short-distance limit of the equations of motion, one concludes that the conditions (\ref{BC_b_zero}) can be fulfilled only if $C=0$. Equation (\ref{Eq_rho}) then implies
\begin{equation}
\rho=\text{const.}=\rho_{vac} \; ,
\end{equation}
or, in terms of the original field variables,
\begin{equation}\label{b_eq}
(-1+6\xi_\chi)\chi^2+(1+6\xi_h)h^2=M_P^{*2} \; ,  
\end{equation}
where
\begin{equation}\label{Mps}
M_P^{*2}=M_P^2\dfrac{-1+6\xi_\chi}{\xi_\chi} \; .
\end{equation}
Thus, in the plane spanned by the vectors $\chi\sqrt{-1+6\xi_\chi}$ and $h\sqrt{1+6\xi_h}$, the bounce trajectory draws an arc of the circle, with the endpoints at $\theta(\infty)=0$ and some $\theta(0)<\theta_{max}$; see Fig. \ref{fig:BI}(a). The radius of the circle is determined by the dilaton vev as the latter appears in the r.h.s. of eq. (\ref{b_eq}). Exploiting the Einstein equations, one obtains the tunneling action:
\begin{equation}\label{B_bounce}
B_{bounce}=-2\pi^2\int_0^\infty dr\:r^3f\tilde{V}(\theta) \; .
\end{equation}
The correction to this formula coming from the non-zero Higgs vev is negligible; see the discussion in sec. \ref{ssec:Remarks}.

Let us now restore the higher-dimensional operator in the dilaton sector. One can expect that this operator provides a small correction to the above solution as long as $\delta\ll 1$. The correction to the tunneling action is given by
\begin{equation}\label{dB_bounce}
\delta B_{bounce}=2\pi^2\delta\int_0^\infty dr\: r^3f^{-3}\theta'^4\tan^4\theta 
\end{equation}
and is small compared to the main contribution (\ref{B_bounce}) even for $\delta=1$; see Fig. \ref{fig:BI2}(b). Still, as we will see, for the ratio of the Higgs vev generated due to instantons to the classical dilaton vev to be exponentially small, one should require $\delta\ll 1$. 

\subsection{The instanton}
\label{ssec:Inst}

According to eqs. (\ref{NewCoord}) and the discussion in sec. \ref{sec:Intro}, the functional whose saddle points are the instanton solutions we are interested in is given by
\begin{equation}\label{Action_Enh_Rho}
\mathcal{B}=-\dfrac{\rho(0)}{M_P}+\int d^4x\L_E \; ,
\end{equation}
where $\L_E$ is given in eq. (\ref{L_E}). The first term is the instantaneous source of the radial field. The source term fixes the center of the configuration and provides an additional boundary condition on it. Together with the vacuum boundary conditions (\ref{BC_Inf}) and the condition for $\theta$ to approach a definite value at $r\rightarrow 0$, they select a unique solution of the Euclidean equations of motion. Due to the source, this solution is singular, hence the name ``singular instanton'' adopted in \cite{Shaposhnikov:2018xkv,Shaposhnikov:2018jag} (see also \cite{Hawking:1998bn}).

Since the dilaton four-derivative operator is suppressed by $M_P^4$, we expect it to affect the instanton behavior only at short distances, $r\lesssim M_P^{-1}$. The tail of the instanton, which contributes to $B_{LE}$, is independent of $\delta$. For this reason, below we consider first the case $\delta=0$, and then the general case $\delta>0$.

\subsubsection{The case $\delta=0$}
\label{sssec:d0}

\begin{figure*}[t]
	\begin{center}
		\begin{minipage}[h]{0.49\linewidth}
			\center{\includegraphics[scale=0.65]{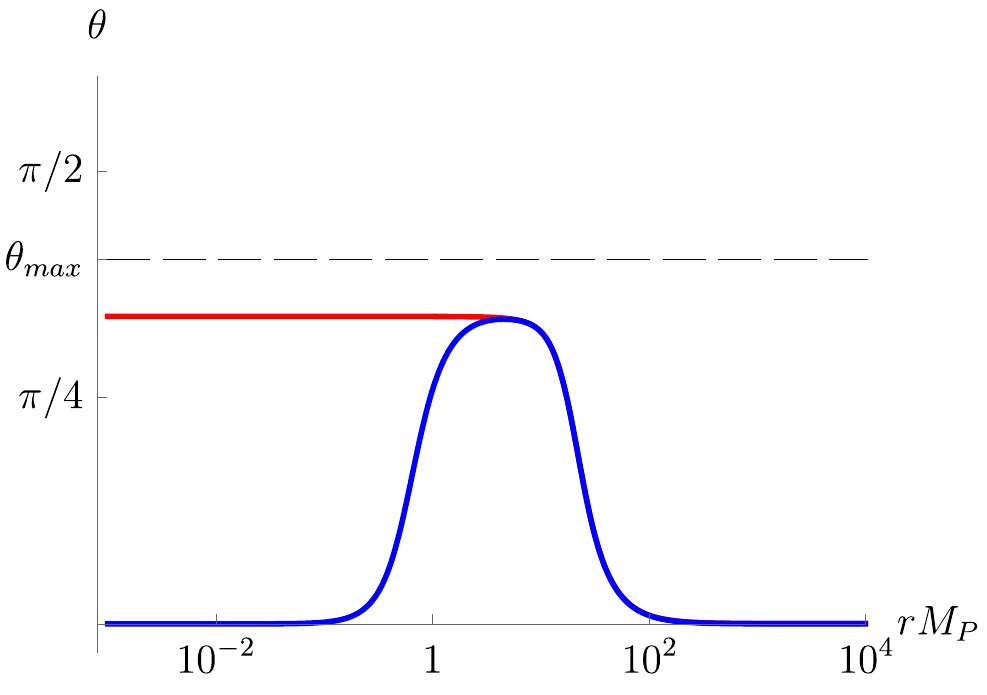} \\ (a)}
		\end{minipage}
		\begin{minipage}[h]{0.49\linewidth}
			\center{\includegraphics[scale=0.65]{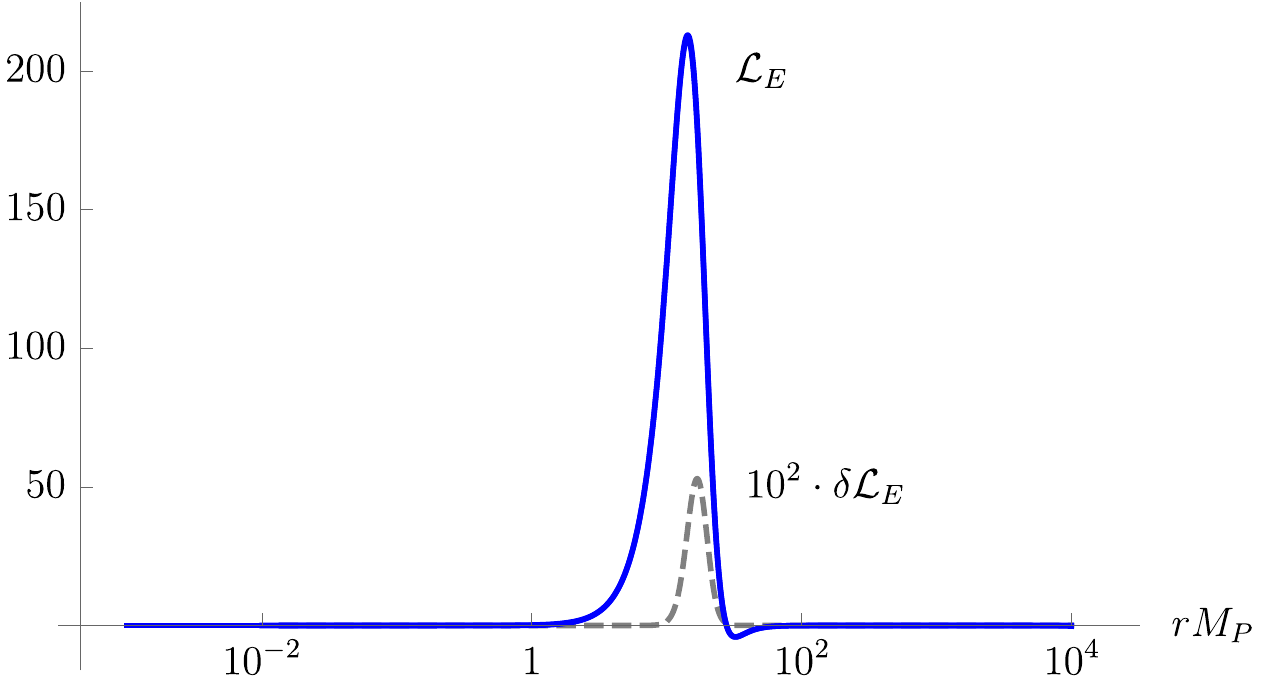} \\ (b)}
		\end{minipage}
		\caption{ \textit{Left panel:} The profile of the angular field component of the bounce (the solid red line with non-zero asymptotics at $r\rightarrow 0$) and of the instanton (the solid blue line with zero asymptotics at $r\rightarrow 0$). One sees that at $rM_P\gg 1$ the two Euclidean solutions almost coincide. The dashed line denotes the upper limit of $\theta$, above which the kinetic term of the scalar fields of the model (\ref{L_J}) is not positive definite. \textit{Right panel:} The solid line is the Euclidean Lagrangian (\ref{L_E}) computed on the bounce or on the instanton. The dashed line is the correction to the euclidean Lagrangian, $\delta\L_E$, due to non-zero $\delta$, $\delta B_{bounce}=\int drM_P \delta \L_E$ (see eq. (\ref{dB_bounce})). Here we take $\delta=1$. The parameters of the model are $\lambda(\theta)=0.01\cos 4\theta$, $\xi_\chi=0.2$, and $\xi_h=0.02$.  }
		\label{fig:BI2}
	\end{center}
\end{figure*}

Varying eq. (\ref{Action_Enh_Rho}) with respect to $\rho$, we arrive at eq. (\ref{Eq_rho}) with
\begin{equation}
C=-\dfrac{1}{M_P} \; .
\end{equation}
Hence, for the instanton $\rho\neq$const. and this makes it differ qualitatively from the bounce, at least near the origin. The analysis of the equations of motion gives the following asymptotics of the radial field in the core region:
\begin{equation}\label{As_rho_zero}
\rho\sim -M_P\gamma\log M_Pr \; , ~~~ \gamma=\sqrt{6a_0} \; , 
\end{equation}
where
\begin{equation}\label{a0}
a_0\equiv a(0)=\dfrac{1}{6-1/\xi_\chi} \; ,
\end{equation}
and $a(\theta)$ is defined in eqs. (\ref{VariousF}). Next, we require the angular field $\theta$ to approach some definite finite value at $r\rightarrow 0$. Otherwise, the solution will experience infinite oscillations which, via the four-derivative term, will produce an infinite contribution to the instanton action. Note that this argument to discard the oscillating solutions does not depend on a particular structure of the model \cite{Shaposhnikov:2018jag}. Then, the only admissible short-distance asymptotics for $\theta$ are
\begin{equation}\label{As_theta_zero}
\theta\rightarrow \pi k \; , ~~~ k=0,1,2,... \; , ~~~ r\rightarrow 0 \; .
\end{equation}
Again, we see the difference from the bounce for which any positive $\theta<\theta_{max}$ is suited. We focus on the case $k=0$, since numerical analysis shows that for all $k>0$ the solutions do not fall off at infinity as required by eqs. (\ref{BC_Inf}). The asymptotic behavior of $\theta$ is 
\begin{equation}\label{As_theta_zero2}
\theta\sim r^\beta \; , ~~~ \beta=\gamma\sqrt{1+\xi_h/\xi_\chi} \; .
\end{equation}

The crucial observation is that the angular field is not a monotonically decreasing function of $r$. Instead, it grows from zero at the center of the instanton up to some finite value and then bounces back and falls down to zero at large $r$; see Fig. \ref{fig:BI}(b). This somewhat surprising behavior is due to the asymptotics (\ref{As_theta_zero}). The latter, in turn, are traced back to the sign of the dilaton kinetic term in eq. (\ref{L_chi}).\footnote{The ``usual'' choice of the sign would result in the asymptotics $\theta\rightarrow\pi /2$ and in the monotonic solution; see sec. 4.3 of \cite{Shaposhnikov:2018jag}.} 

The turning point at which the bounce occurs specifies the second characteristic scale of the instanton solution, and this scale can be many orders of magnitude smaller than the Planck scale determining the size of the core region. The behavior of the instanton at this scale is controlled not by the higher-dimensional Planck-suppressed operators but by the effective Higgs potential $\tilde{V}(\theta)$. It turns out that the turning point (and, hence, the instanton itself) exists under the same conditions as the tunneling solution considered in sec. \ref{ssec:Bounce}. As Fig. \ref{fig:BI2} demonstrates, near the turning point the instanton profile is indistinguishable from the profile of the tunneling solution. If the energy scale probed by the bounce is much below $M_P$, the contribution of the tail of the instanton to the action is nearly the same as the contribution of the bounce:
\begin{equation}\label{B_LE}
B_{LE}\approx B_{bounce} \; .
\end{equation}
In the computations below the two characteristic scales of the instanton will be well separated, and eq. (\ref{B_LE}) will hold with great accuracy.

It is worth mentioning again that our considerations are limited to the range $0\leqslant\theta<\theta_{max}$, where $\theta_{max}$ is defined in eqs. (\ref{PosBounds}). The turning point of the bounce and of the instanton is determined by the shape of the potential $\tilde{V}(\theta)$; it may be above $\theta_{max}$, in which case the model (\ref{L_J}) must be extended. However, having the potential fixed, it is possible to adjust the parameters of the model (mainly, the non-minimal coupling of the Higgs field $\xi_h$) so that the solution does not go beyond the threshold, and we follow this strategy.

Shown in Fig. \ref{fig:da}(a) is the dependence of $B_{LE}$ on the non-minimal couplings $\xi_\chi$, $\xi_h$ and for some choice of the function $\lambda(\theta)$. We see that varying the Higgs non-minimal coupling $\xi_h$ does not change $B_{LE}$, at least when $\xi_h\lesssim 10^{-2}$. The range of $\xi_h$ is confined to small values to assure that the solution stays within the region of validity of Lagrangian (\ref{L_J}).

Let us now turn to the high-energy part of the instanton, where the fields follow the asymptotics (\ref{As_rho_zero}) and (\ref{As_theta_zero2}). From eqs. (\ref{Action_Enh_Rho}) and (\ref{As_rho_zero}) we see immediately that the divergence of the radial field at the origin prevents the finite contribution to $B_{HE}$. This issue is resolved by switching on the four-derivative operator, which makes the magnitude of $\rho$ at the center of the instanton finite, thus removing the divergence and partially regularizing the singularity. We proceed to this step below.

\subsubsection{The case $\delta\neq 0$}
\label{sssec:d1}

Expressing $\chi$ via $\rho$ and $\theta$ in the four-derivative operator, one obtains
\begin{equation}\label{Op_d}
\sqrt{g}\delta\dfrac{(\d\chi)^4}{\chi^4} = \sqrt{\tilde{g}}\delta\dfrac{(\tilde{\d}\rho)^4}{M_P^4} +...
\end{equation}
where dots stand for the terms proportional to $\sin\theta\cdot\d\theta$. At large distances from the center of the instanton, these terms are negligible, the argument being the same as in the case of the bounce. At short distances these terms are again negligible due to the asymptotics (\ref{As_theta_zero2}). Thus, the only sizable effect from introducing the higher-dimensional operator in the dilaton sector is the modification of the equation for the radial field:
\begin{equation}\label{Eq_rho_mod}
\dfrac{4\delta}{M_P^4}\dfrac{\rho'^3r^3}{f^3}+\dfrac{\rho'r^3}{fa(\theta)}=-\dfrac{1}{M_P} \; .
\end{equation}
The first term in this equation becomes dominant when $r\lesssim\bar{r}$, where $\bar{r}$ is found from
\begin{equation}
\bar{r}=M_P^{-1}\delta^{1/6}a(\theta(\bar{r}))^{1/2} \; .
\end{equation}
Inspecting the shape of the instanton, one concludes that, in fact, $\a(\theta(\bar{r}))\approx a_0$; hence
\begin{equation}\label{RBar}
\bar{r}\approx M_P^{-1}\delta^{1/6}a_0^{1/2} \; .
\end{equation}
Solving the equations of motion at $r\lesssim\bar{r}$, one obtains
\begin{equation}\label{As_rho_zero_d}
\rho'\sim -M_P^2\delta^{-1/6} \; .
\end{equation}
We see that the radial field does not diverge anymore. The higher-dimensional operator (\ref{Op_d}) partially cures the singularity of the solution. The cure is not perfect, as one can see, e.g., by computing the scalar curvature near the origin:
\begin{equation}
\tilde{R}\sim r^{-2} \; , ~~~ r\lesssim\bar{r} \; .
\end{equation}
It is, however, enough for our purposes, as the contribution to the instanton action (\ref{Action_Enh_Rho}) supplemented with the source term is now finite. Indeed, for the source term one can make the following estimation \cite{Shaposhnikov:2018jag}:
\begin{equation}\label{Rho_0}
\rho(0)/M_P\sim a_0^{1/2}(\log(\delta a_0^{-3})+\mathcal{O}(1)) \; .
\end{equation}
Next, using the Einstein equations and eq. (\ref{As_rho_zero_d}), one obtains, for the high-energy contribution to the instanton action,
\begin{equation}\label{B_d}
2\pi^2\int_0^{\bar{r}}dr\:r^3f\delta\dfrac{\rho'^4}{f^4M_P^4}\sim a_0^{1/2} \; .
\end{equation}
Overall, $B_{HE}$ is finite. It is interesting to note that, despite being singular and, in the case $\delta=0$, even divergent, the instanton brings a finite contribution to the Euclidean action \cite{Hawking:1998bn}. Note also that, somewhat unexpectedly, eq. (\ref{B_d}) shows no power-like dependence of the instanton action on $\delta$.

\begin{figure*}[t]
	\begin{center}
		\begin{minipage}[h]{0.49\linewidth}
			\center{\includegraphics[scale=0.65]{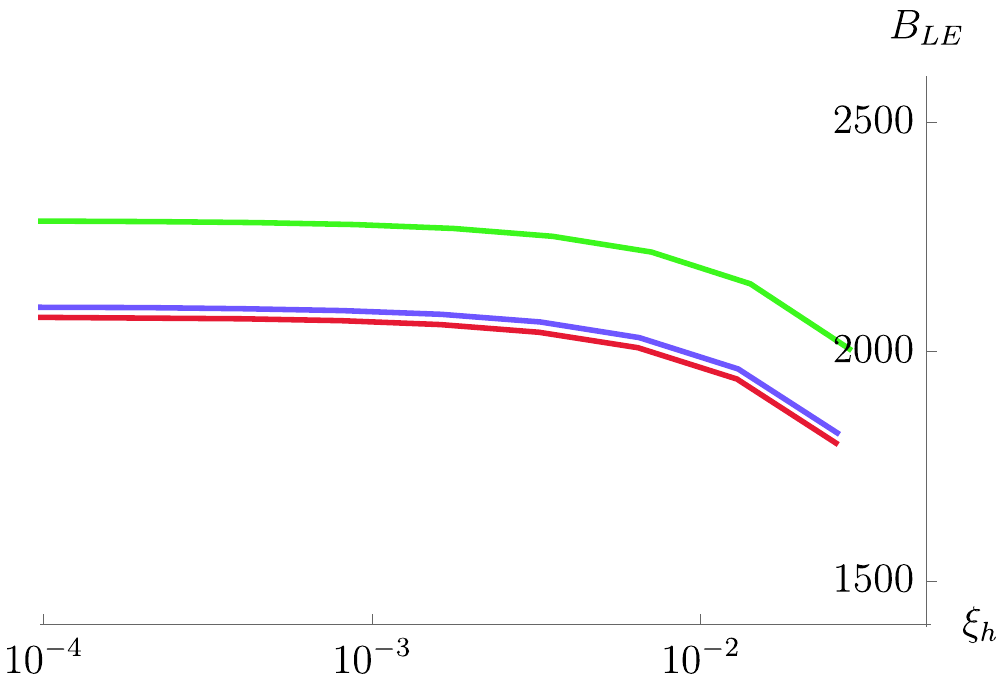} \\ (a)}
		\end{minipage}
		\begin{minipage}[h]{0.49\linewidth}
			\center{\includegraphics[scale=0.65]{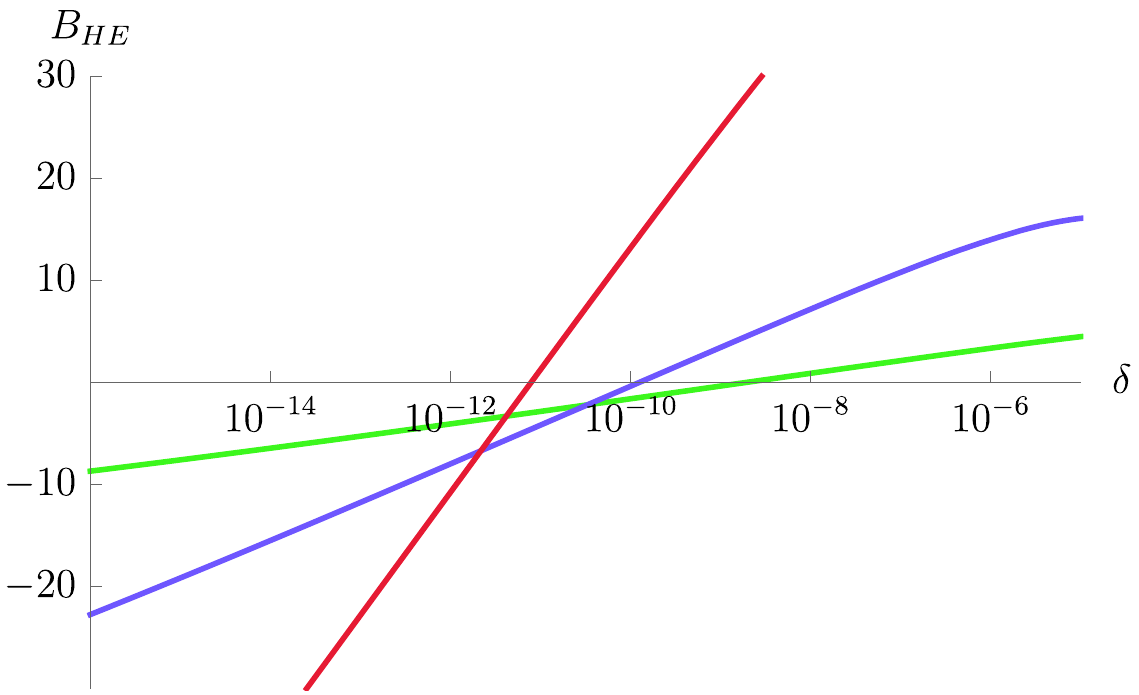} \\ (b)}
		\end{minipage}
		\caption{ \textit{Left panel:} The low-energy contribution to the instanton action (\ref{Action_Enh_Rho}) as a function of $\xi_h$ and for different values of $\xi_\chi$. From the top to the bottom line, $6\xi_\chi-1=10^{-1}$, $10^{-2}$, $10^{-3}$. We see that $B_{LE}$ is independent of $\xi_h$, at least when $\xi_h\lesssim 10^{-2}$. Here $\lambda(\theta)=0.01\cos 4\theta$. \textit{Right panel:} The high-energy contribution to the instanton action (\ref{Action_Enh_Rho}) as a function of $\delta$ and for the different values of $\xi_\chi$. From the low to the large inclination, $6\xi_\chi-1=10^{-1}$, $10^{-2}$, $10^{-3}$. We see that $|B_{HE}|$ shows power-like dependence on the deviation of $\xi_\chi$ from the conformal limit, in accordance with eq. (\ref{B_HE}), while the sign of $B_{HE}$ is different for the different values of $\delta$, and small $\delta$ are needed to obtain $B_{HE}<0$.   }
		\label{fig:da}
	\end{center}
\end{figure*}

From eqs. (\ref{Rho_0}) and (\ref{B_d}) we conclude that
\begin{equation}\label{B_HE}
B_{HE}\sim a_0^{1/2} \; .
\end{equation}
This means that, unless a cancellation occurs between the terms with sub-dominant dependence on $a_0$, the high-energy part of the instanton action (\ref{Action_Enh_Rho}) can be made large by making the coefficient $a_0$ large. According to eq. (\ref{a0}), large values of $a_0$ are achieved by choosing the dilaton non-minimal coupling $\xi_\chi$ to be close to the conformal limit, $6\xi_\chi-1\ll 1$. We will say more about this in sec. \ref{sec:DiscConcl}.\footnote{ The condition $a_0\gg 1$ can also be understood from a geometric point of view. From eqs. (\ref{L_E}) and (\ref{a0}) one can infer that sending $a_0$ to infinity amounts to blowing up the curvature in the space of scalar fields $\rho$, $\theta$; see \cite{Karananas:2016kyt}.   } The quantity $a_0^{-1}$ serves  as a small parameter that justifies the applicability of the saddle-point approximation made in obtaining eq. (\ref{Hierarchy}).

The simple estimate (\ref{B_HE}) does not allow us to compute the sign of $B_{HE}$. The latter is determined by the balance between the negative source term and the positive contribution from the higher-dimensional operator. As Fig. \ref{fig:da}(b) shows, both signs are realized for different choices of the parameters $a_0$ and $\delta$. Recall that our goal is to make the total contribution $B_{HE}+B_{LE}$ of the order $10$, in which case the hierarchy (\ref{TheHierarchy}) between the weak and the Planck scales is reproduced. In view of eq. (\ref{B_LE}), this means that one should look for the values of $a_0$ and $\delta$ for which $B_{HE}<0$ and $|B_{HE}|=\mathcal{O}(10^3)$. In the next section we will see that this puts strong constraints on the parameters of the model. Furthermore, this implies some fine-tuning between the Higgs and the dilaton sectors of Lagrangian (\ref{L_J}).

\section{The hierarchy}
\label{sec:Hierarchy}

In the previous section we analyzed the properties of the instanton solution in the model (\ref{L_J}) with the toy potential for the field $h$. Our conclusion was that the total instanton action $B$ consists of the equally important contributions from the core of the instanton, $B_{HE}$, and from the tail of the instanton, $B_{LE}\approx B_{bounce}$. The former can be of any magnitude and sign, depending on the parameters of the model, mainly on the dilaton non-minimal coupling $\xi_\chi$ and the quartic derivative coupling $\delta$. It is, therefore, possible, by adjusting these parameters, to obtain $B=\log M_P/v$. 

Let us embed the model (\ref{L_J}) into a realistic setting. First, we identify $h$ with the Higgs field degree of freedom in the unitary gauge,
\begin{equation}
\varphi=1/\sqrt{2}\;(0,h)^T \; .
\end{equation}
Then, we supplement Lagrangian (\ref{L_J}) with the rest of the SM content. This content does not modify the results of the leading-order saddle-point evaluation of the action (\ref{Action_Enh_Rho}). The SM fields are the source of perturbative corrections to the Higgs vev. These corrections are well known \cite{Coleman:1973jx} and, as was mentioned in sec. \ref{sec:Intro}, they are not capable of generating the observed value of the EW scale. Note also that, since the dilaton sector is decoupled from the rest of the theory, the dilaton field contributes to $\langle h\rangle$ only via graviton loops, and this contribution is numerically small at low energies \cite{tHooft:1974toh,Bezrukov:2012hx}. Thus, under the assumption of the absence of new heavy particles coupled to the Higgs field, the instanton studied above provides the dominant, non-perturbative contribution to $\langle h\rangle$.

\begin{figure*}[t]
	\begin{center}
		\begin{minipage}[h]{0.49\linewidth}
			\center{\includegraphics[scale=0.65]{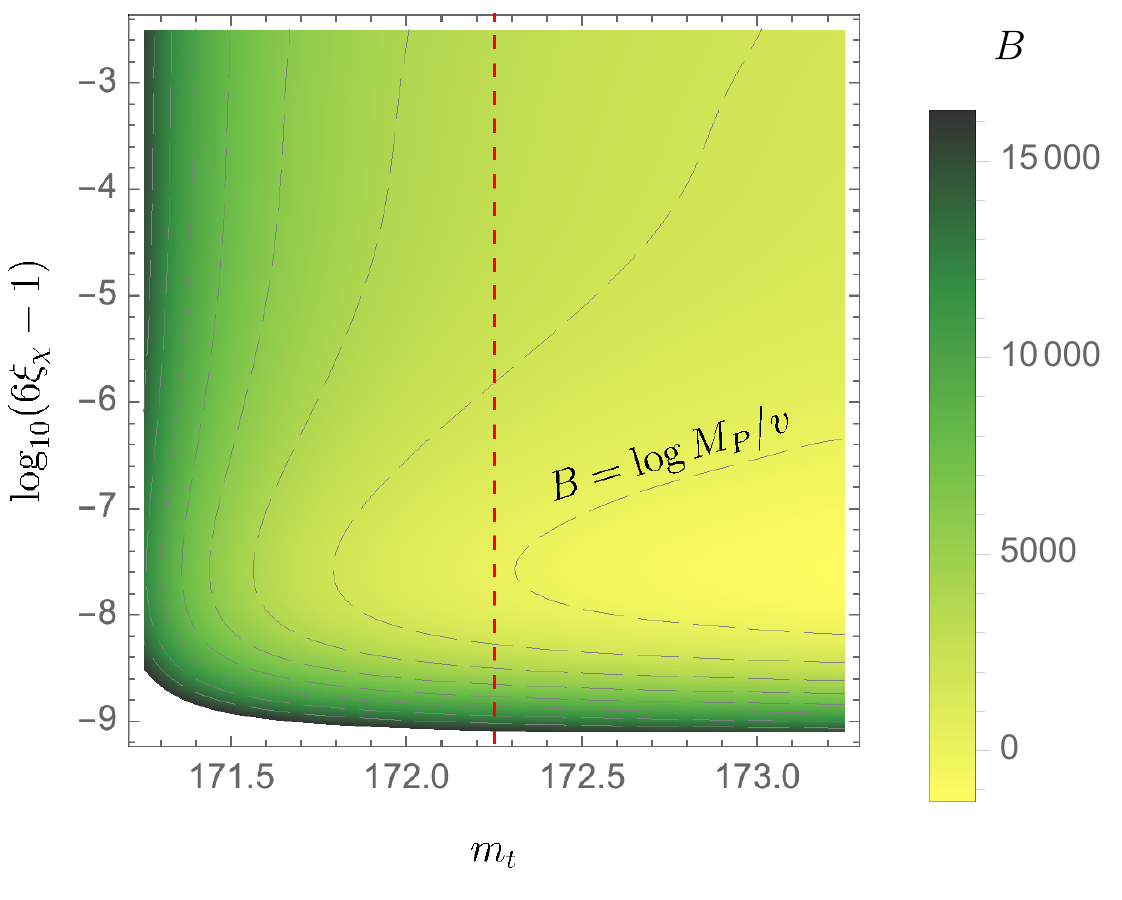}}
		\end{minipage}
		\begin{minipage}[h]{0.49\linewidth}
			\center{\includegraphics[scale=0.65]{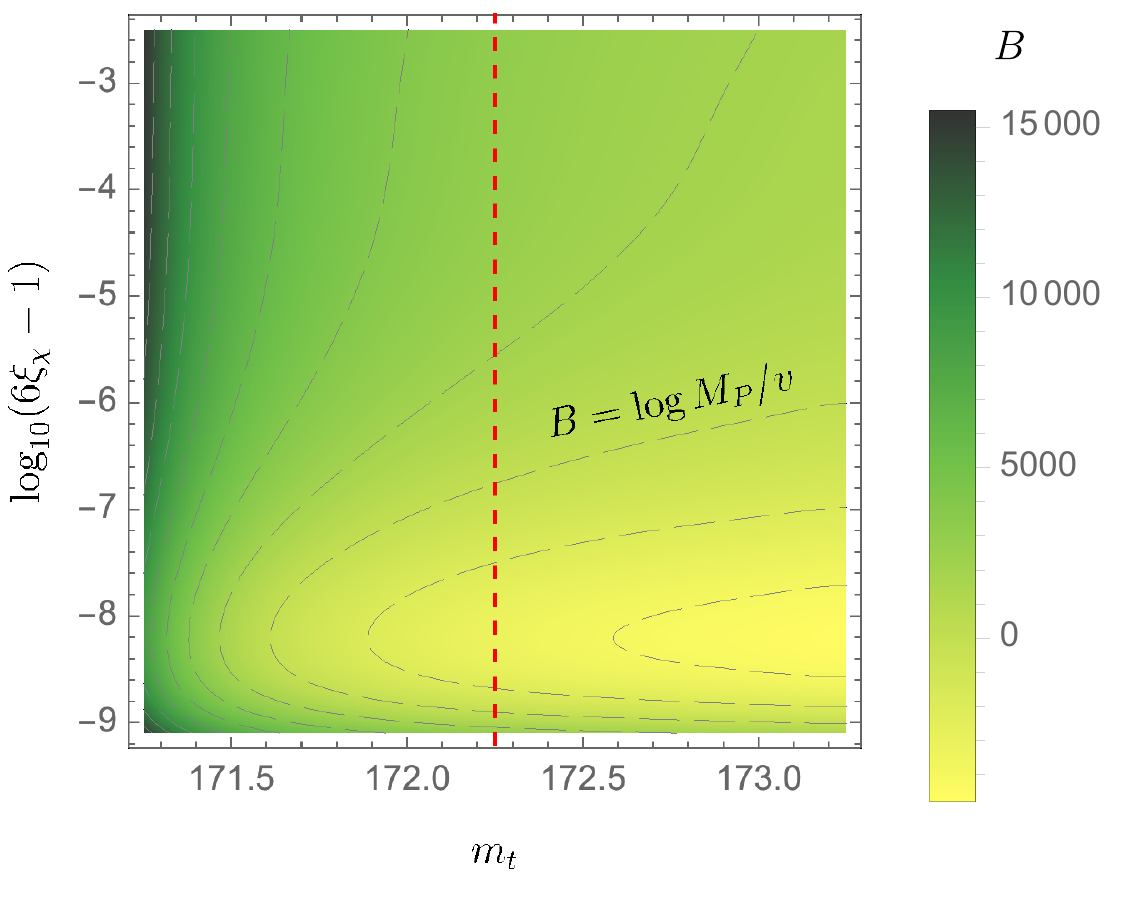}}
		\end{minipage}
		\caption{ The instanton action $B$ plotted against $m_t$ and $\xi_\chi$, with $\xi_h=0.02$ and $\delta=10^{-18}$ (left panel), $\delta=10^{-20}$ (right panel). The grey dashed isoparametric curves are plotted with steps $\Delta B=\pm 2000$, starting from $B=\log M_P/v\approx 37$. Here we take the function $\lambda(\theta)$ as the Higgs self-coupling constant undergoing RG running within the SM and under the SI renormalization scheme; see the text for details. The vertical dashed line marks the central value of the top mass $m_t=172.25$ GeV \cite{Castro:2017yxe}, and the running of $\lambda$ is computed at the central value of the Higgs mass $m_H=125.09$ GeV \cite{Aad:2015zhl}. }
		\label{fig:Final}
	\end{center}
\end{figure*}

Although the fields other than $\chi$, $h$ and $g_{\mu\nu}$ do not participate in building the instanton solution, their presence is important because of their interaction with the Higgs field. The interaction determines the RG running of the Higgs quartic coupling. With the aim to explicitly preserve the scale symmetry of the model at the perturbative quantum level, we implement the SI renormalization procedure \cite{Englert:1976ep,Shaposhnikov:2008xi,Tamarit:2013vda,Ghilencea:2016ckm,Ghilencea:2016dsl,Mooij:2018hew}. It amounts to replacing the 't Hooft-Veltman normalization point $\mu$ \cite{tHooft:1972tcz} by a field-dependent scale, $\mu^2=F(h,\chi)\hat{\mu}^2/M_P^2$, where the function $F$ reflects the choice of the renormalization prescription and $\hat{\mu}$ denotes the usual momentum scale in the RG equations, $\lambda=\lambda(\hat{\mu})$, on which nothing depends in the final result.\footnote{Note that, since the theory under consideration is non-renormalizable, the results depend on the choice of the function $F$ \cite{Shaposhnikov:2009nk,Bezrukov:2012hx,Shaposhnikov:2018nnm}. } We choose the momentum scale as follows \cite{Bezrukov:2009db,Bezrukov:2012hx}:
\begin{equation}
\hat{\mu}^2=\dfrac{y_t^2}{2}\dfrac{M_P^2h^2}{\xi_\chi\chi^2+\xi_hh^2} \; ,
\end{equation}
where $y_t$ is the top quark Yukawa coupling. In the polar field variables, this gives
\begin{equation}\label{MuTheta}
\hat{\mu}^2=\dfrac{y_t^2}{2\xi_h}\dfrac{M_P^2}{1+\zeta\cot^2\theta} \; ,
\end{equation}
and, according to eqs. (\ref{SI_E}), the scale symmetry is manifestly preserved.

Currently, the largest uncertainty in the shape of the effective Higgs potential comes from uncertainties in the top mass measurements \cite{Bezrukov:2014ina,Spannagel:2016cqt,YauWong:2016idk}. Therefore, we use $m_t$ as the parameter controlling the low-energy contribution to the instanton action. Note that it is possible, within the current uncertainties, for $\lambda$ to stay positive all the way up to the Planck scale \cite{Bezrukov:2014ina}, in which case no instanton with the finite action exists in the model (\ref{L_J}). Thus, the mechanism of generating the hierarchy of scales discussed here strongly relies on the metastability of the EW vacuum. The RG running of $\lambda$ is computed using the code based on \cite{Chetyrkin:2012rz,Bezrukov:2012sa}. 

Shown in Fig. \ref{fig:Final} are dependencies of the total instanton action $B$ on $m_t$ and $\xi_\chi$, for a fixed $\xi_h=0.02$ and for the two values of $\delta$. The range of $m_t$ is chosen around the central value $m_t=172.25$ GeV \cite{Castro:2017yxe}. We take $m_t$ to be above the critical value separating the domains of stability and metastability of the EW vacuum. The value of $\xi_h$ is chosen to stay within the region of validity of Lagrangian (\ref{L_J}). As was discussed in sec. \ref{sec:BI}, the dependence of $B$ on $\xi_h$ is very mild, at least for $\xi_h\lesssim 10^{-2}$; see Fig. \ref{fig:da}(a).

\begin{figure}[b]
\begin{center}
		\begin{minipage}[h]{0.45\linewidth}
			\center{\includegraphics[scale=0.22]{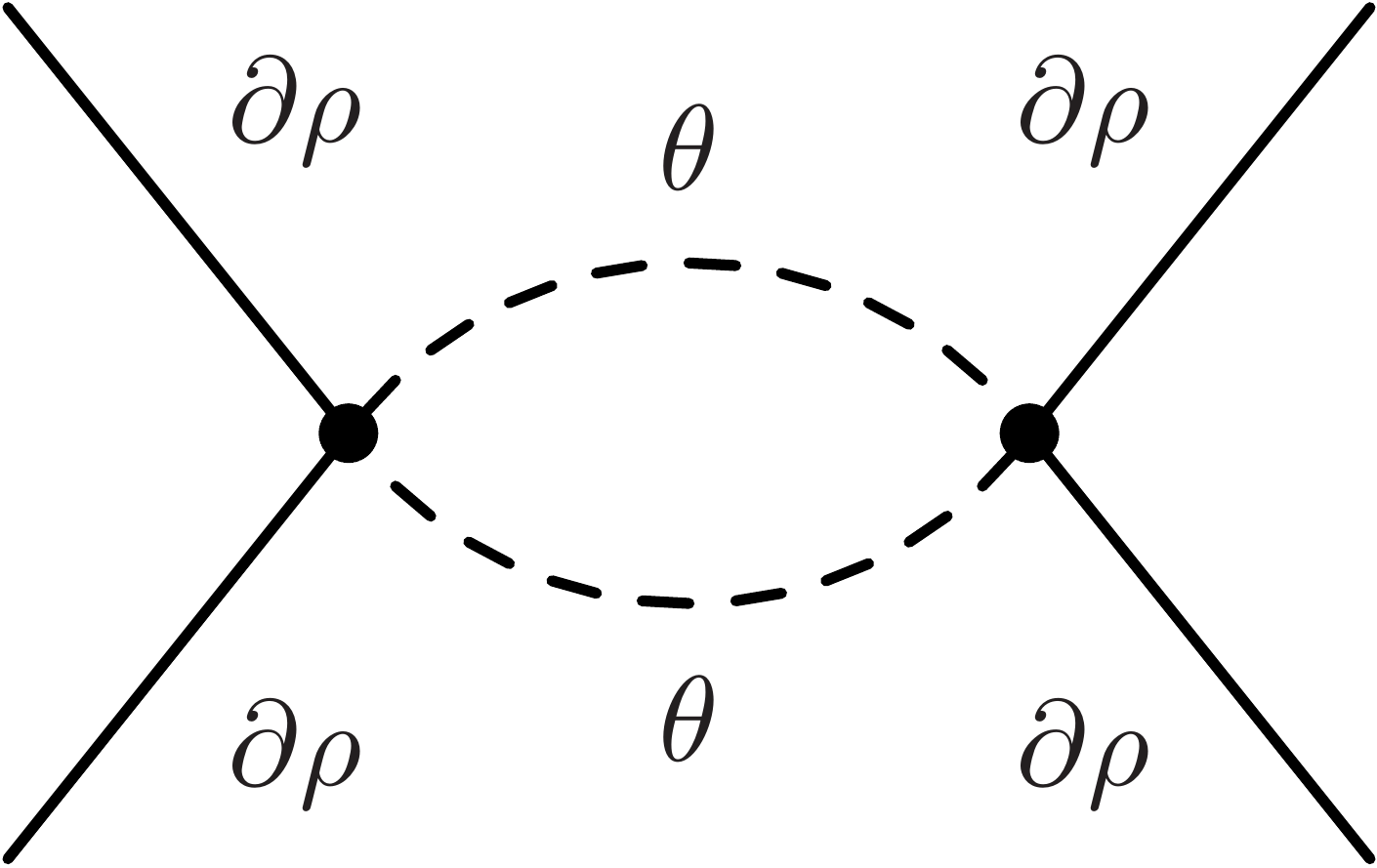}}
		\end{minipage}
		\begin{minipage}[h]{0.45\linewidth}
			\center{\includegraphics[scale=0.22]{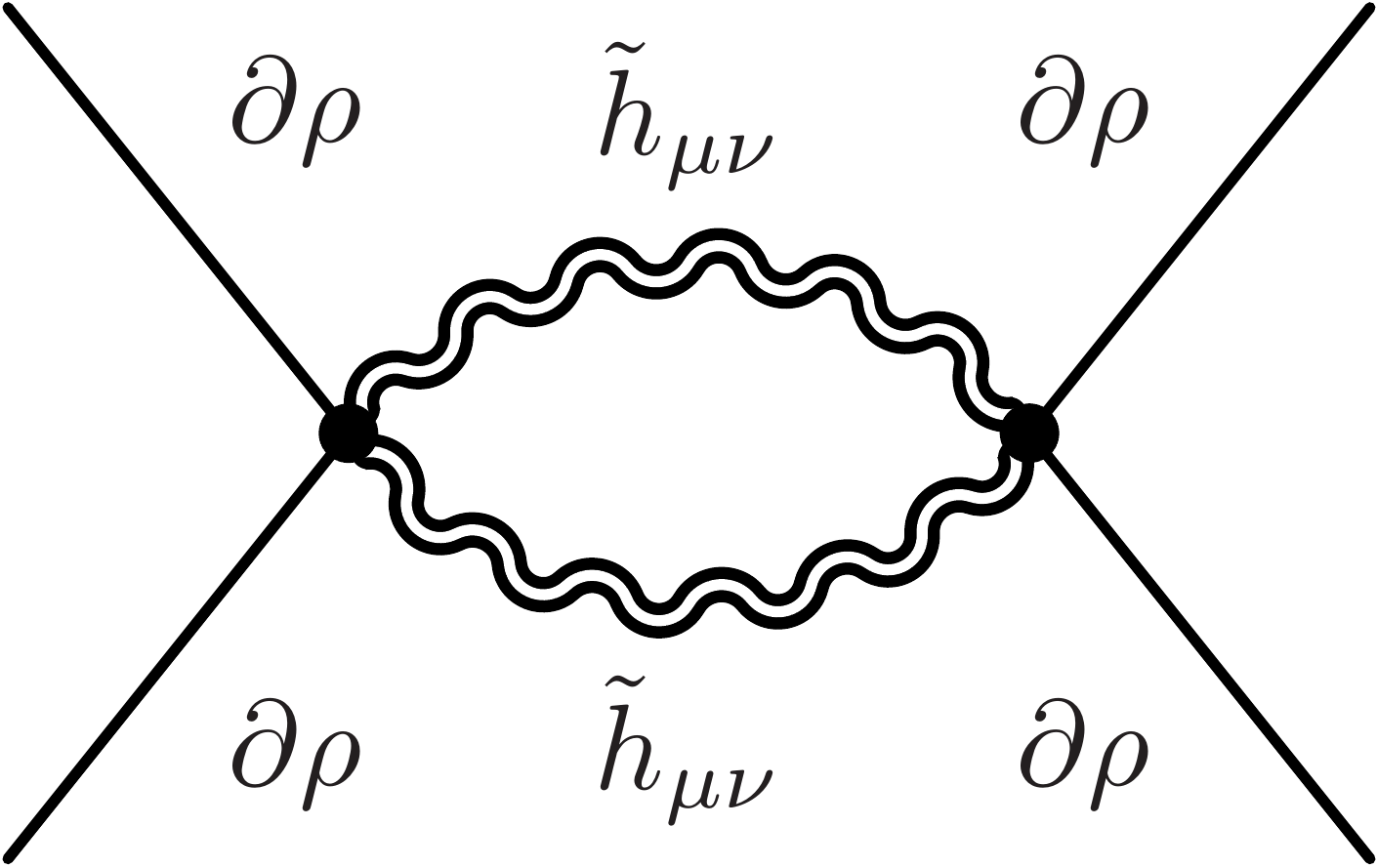}}
		\end{minipage}
	\caption{Schematic form of diagrams providing the leading correction to $\delta$ in perturbation theory above the classical vacuum (\ref{GroundState_J}). Here $\tilde{h}_{\mu\nu}$ is the metric fluctuation above the flat background. }
	\label{fig:Diagram}
\end{center}
\end{figure}

We observe that the values $B=\mathcal{O}(10)$ are achieved in a certain window of the parameter space. Namely, the dilaton coupling must be close to the conformal limit, $6\xi_\chi-1 \leq \mathcal{O}(10^{-5})$. As for the coupling $\delta$, it must be extremely close to zero, $0<\delta\lesssim 10^{-16}$, at least when $m_t$ stays within the $2\sigma$ uncertainty region $172.25\pm 1.26$ GeV \cite{Castro:2017yxe}. This constraint is due to the requirement for $B_{HE}$ to be large and negative, in order to compensate the large and positive contribution from the potential; see Fig. \ref{fig:da}(b). It is also due to the fact that $\delta$ enters the expression for $B_{HE}$ only logarithmically. Despite its smallness, the coupling does not affect much the energy scale at which the UV part of the model comes into play, as it follows, e.g., from eq. (\ref{RBar}). Note also that $\delta$ does not bring about new interaction scales much below $M_P$. Moreover, having small $\delta$ is natural, since setting it to zero (along with taking the limit $\xi_\chi\rightarrow 1/6$) enhances the symmetry of the dilaton sector by making it Weyl invariant. One can check that at low energies no fine-tuning is required to keep $\delta$ that small. Indeed, in perturbation theory the main correction to $\delta$ comes from Higgs and graviton loops; see Fig. \ref{fig:Diagram}.
They generate a correction of the order of $ (\xi_\chi-1/6)^2=(36a_0)^{-2}$. Thus, as soon as $\delta \gtrsim (\xi_\chi-1/6)^2$, no fine-tuning is needed to set it as close to zero as necessary. From Fig. \ref{fig:Final} we see that this condition is consistent both with having $B=\log M_P/v$ and with the top mass lying in its current uncertainty region close to the central value.

\section{Discussion and conclusion}
\label{sec:DiscConcl}

Let us summarize our findings and outline the features of the non-perturbative mechanism of generation of a new scale, the ones which are specific to the model (\ref{L_J}), and the ones which this model has in common with the theories studied previously in \cite{Shaposhnikov:2018xkv,Shaposhnikov:2018jag}.

First of all, the non-minimal couplings of the scalar fields to gravity are the essential ingredient of the mechanism. When evaluating the vev of the scalar field, these couplings naturally lead to the appearance of the source term. This happens because the low-energy canonical scalar degrees of freedom are related to their high-energy counterparts via an exponential mapping.

The higher-dimensional operator (\ref{Op_d}) is also an important ingredient of the mechanism. Thanks to this operator, the instanton configuration does not diverge at the origin, leading to the finite contribution from the source term. As for the instanton action, it is finite even without the partial regularization provided by the derivative operator.\footnote{This fact was employed in \cite{Hawking:1998bn}, where similar singular classical configurations were treated as tunneling solutions. } Note that, unlike the theories studied in \cite{Shaposhnikov:2018jag}, in our case it is sufficient to consider UV modifications of the dilaton sector, while leaving the Higgs sector intact. This is the consequence of the peculiar behavior of the instanton solution in the model (\ref{L_J}). Since the dilaton field is coupled to the rest of the theory only gravitationally, probing different UV operators composed of $\chi$ does not endanger the SM physics. The operator (\ref{Op_d}) provides us with the simplest but not unique possibility to cure the divergence of the instanton. Other types of SI operators containing $\chi$ and its first derivative can also work; see the discussion in \cite{Shaposhnikov:2018jag}.

As was already pointed out, the main feature of the instanton contribution to the Higgs vev in the model (\ref{L_J}) is that it consists of two equally important parts. One part is saturated in the core of the instanton, where the higher-dimensional operator from the dilaton sector takes over, and the other is saturated in the tail where the physics much below the Planck scale dominates. As we see, the necessary instanton existence condition is the negativity of $\lambda$ in a certain region of energy scales, which implies metastability of the Higgs vacuum, given that $B_{bounce}=\mathcal{O}(10^3)$. This enables us to constrain the mechanism from experiment.

Equation (\ref{B_LE}) tells us that the high-energy contribution to the instanton action must compensate two orders of magnitude in the difference between $B_{LE}$ and the desired value of $B$. This implies a balance between different energy domains of the theory and between its dilaton and Higgs sectors. This balance is successful for quite special values of the parameters of the dilaton sector, $\xi_\chi$ and $\delta$. Namely, they must be close to the point in the parameter space where the dilaton Lagrangian (\ref{L_chi}) becomes Weyl invariant. We find this interesting, since in the theories considered in \cite{Shaposhnikov:2018xkv,Shaposhnikov:2018jag}, the approximate Weyl invariance in the high-energy regime was also identified as one of the conditions necessary for the successful implementation of the mechanism.

Let us make one more comment regarding the dilaton derivative coupling $\delta$. We saw above that it should belong to the interval $0<\delta \lesssim 10^{-16}$, where it is natural and can be made stable against loop corrections at low energies. As was discussed in sec. \ref{ssec:Lagr}, the simplest way to obtain the SM Higgs potential at low energies is to supplement the model (\ref{L_J}) with the Higgs-dilaton coupling term $\alpha\chi^2h^2$. Then, phenomenology requires $\alpha$ to be of the order of $\xi_\chi(m_H/M_P)^2\sim 10^{-33}$ \cite{GarciaBellido:2011de}. One may wonder if our considerations amount to just trading one small coupling ($\alpha$) for another ($\delta$). We believe that this is not the case. The difference between the two ways to obtain the hierarchy of scales is revealed in their respect to the scale symmetry. Indeed, the instanton breaks the scale invariance of the model semiclassically, while the $\chi^2h^2$-term breaks it spontaneously via the vev of the dilaton field. One can contemplate the mechanism that would forbid the Higgs-dilaton coupling term but allow for the dilaton four-derivative term (see, e.g., \cite{Karananas:2016grc}).

To conclude, in this paper we elaborated on the hypothesis that the exponentially small ratio of the Fermi to the Planck scales can be generated by the non-perturbative gravitational effect. At the semiclassical level, the effect manifests itself in the existence of an instanton configuration contributing to the vev of the scalar field. The resulting hierarchy between the scales is determined by the instanton action. Being non-perturbative, this instanton mechanism is sensitive to physics at energy scales as high as the Planck scale. In this paper we demonstrated that the mechanism can also be sensitive to physics at low energies. We worked in the CI SM framework extended by the non-minimal coupling of the Higgs field to gravity and by the dilaton sector, eq. (\ref{L_chi}). The high-energy contribution to the instanton action is controlled by the UV structure of the dilaton sector, while the low-energy contribution is determined by the SM parameters. The sensitivity of the mechanism to the SM physics is a distinct feature of the theory under consideration, as it was absent in the variety of models studied before in \cite{Shaposhnikov:2018xkv,Shaposhnikov:2018jag}. Nevertheless, we found many similarities between all examples that have been considered. Among them are the non-minimal coupling of the scalar fields to gravity and the (approximate) Weyl invariance of the dilaton sector in the UV regime. 

There remain open questions concerning the instanton mechanism and its implementation. One of them concerns fluctuations above the instanton background. Studying the latter may shed some light on physical implications of the singular Euclidean solutions that we use. Another natural question is whether the instantons of a similar kind can be helpful in resolving the cosmological constant problem. So far it is only clear that more or less straightforward attempts to implement the same approach to the vev of the curvature fail. We will address these questions in future work.

\section*{Acknowledgements}

The author thanks Mikhail Shaposhnikov and Kengo Shimada for useful discussions. The work was supported by the Swiss National Science Foundation.


\bibliographystyle{elsarticle-num}
\bibliography{D_gravity}

\end{document}